%% file: main.tex
\documentclass[%
 superscriptaddress,
 reprint,
 amsmath,
 amssymb,
 prl,
 showpacs,
 twocolumn, 
 floatfix,
 nofootinbib,
]{revtex4-2}

\usepackage{graphicx}
\usepackage{float}
\usepackage{dcolumn}
\usepackage{bm}
\usepackage[colorlinks=true, allcolors=blue]{hyperref}
\usepackage[mathlines]{lineno}
\usepackage{color}
\usepackage[normalem]{ulem}
\usepackage{booktabs}

\begin{document}

\title[Two-Stage Cryogenic HEMT Based Amplifier For Low Temperature Detectors]{Two-Stage Cryogenic HEMT Based Amplifier For Low Temperature Detectors}

\input{authors.tex}


\begin{abstract}
To search for dark matter candidates with masses below $\mathcal{O}$(MeV), the SPLENDOR (Search for Particles of Light dark mattEr with Narrow-gap semiconDuctORs) experiment is developing novel narrow-bandgap semiconductors with electronic bandgaps on the order of 1-100 meV. In order to detect the charge signal produced by scattering or absorption events, SPLENDOR has designed a two-stage cryogenic HEMT-based amplifier with an estimated charge resolution approaching the single-electron level. A low-capacitance ($\sim$1.6 pF) HEMT is used as a buffer stage at $T=10\,\mathrm{mK}$ to mitigate effects of stray capacitance at the input. The buffered signal is then amplified by a higher-capacitance ($\sim$200 pF) HEMT amplifier stage at $T=4\,\mathrm{K}$. Importantly, the design of this amplifier makes it usable with any insulating material - allowing for rapid prototyping of a variety of novel detector materials. We present the two-stage cryogenic amplifier design, preliminary voltage noise performance, and estimated charge resolution of 7.2 electrons.
\end{abstract}

\maketitle

\section{Introduction}
Understanding the particle nature of dark matter, which makes up approximately 85\% of the matter content in the universe, remains one of the biggest open questions in the fields of particle physics and cosmology. After decades of null results in searches for weakly interacting massive dark matter candidates~\cite{xenon1t_wimp, PhysRevLett.118.021303, PhysRevLett.119.181302, Alkhatib_2021}, experimental and theoretical efforts have shifted towards a broad range of masses, including lighter mass dark matter candidates with masses below $\mathcal{O}$(MeV)~\cite{battaglieri2017cosmic}. These light mass dark matter particles present a substantial detection challenge, as their relatively low kinetic energy limits the energy deposited in a target to be sub-eV. The SPLENDOR collaboration (Search for Particles of Light dark mattEr with Narrow-gap semiconDuctORs) is developing novel narrow-bandgap single-crystal semiconductors with electronic bandgaps on the order 1-100 meV~\cite{rosa2020colossal, Piva_2021} in order to probe fermionic (bosonic) dark matter masses below an MeV (eV).

When dark matter interacts with a semiconductor, electrons (holes) are excited into the conduction (valence) band. This created charge can be drifted in an electric field and the image current integrated to measure the induced signal. To read out this charge signal, the SPLENDOR collaboration designed and built a two-stage cryogenic charge amplifier utilizing high electron mobility transistors (HEMTs) using similar HEMT transistors that have been used in experiments such as SuperCDMS~\cite{PHIPPS2019181} and RICOCHET~\cite{augier2023demonstration}. This amplifier input stage operates at $\sim$10\,mK to both reduce thermal noise and parasitic capacitance. 

Sub-electron charge resolution has currently been achieved with two different technologies: the Skipper-CCD device~\cite{Fern_ndez_Moroni_2012}, and indirectly through the Neganov–Trofimov–Luke effect in the SuperCDMS HVeV devices~\cite{Ponce_2020, Ren_2021, Agnese_2018}. However, both of these technologies require advanced fabrication techniques and have only been demonstrated on silicon. For the SPLENDOR experiment we are developing a charge readout with $\mathcal{O}(1)$ electron resolution that relies only on a capacitive coupling to the detector material. In this way, the intrinsic charge resolution of the readout should be largely independent of the detector material---allowing for rapid prototyping of different semiconducting detector geometries and materials. While we note that different detector materials may include their own backgrounds (e.g. radio-impurities, dark rates, etc), there should be little difference in the fundamental intrinsic noise of the readout of the materials.

\begin{figure}[h]
    \centering
    \includegraphics[width=\linewidth]{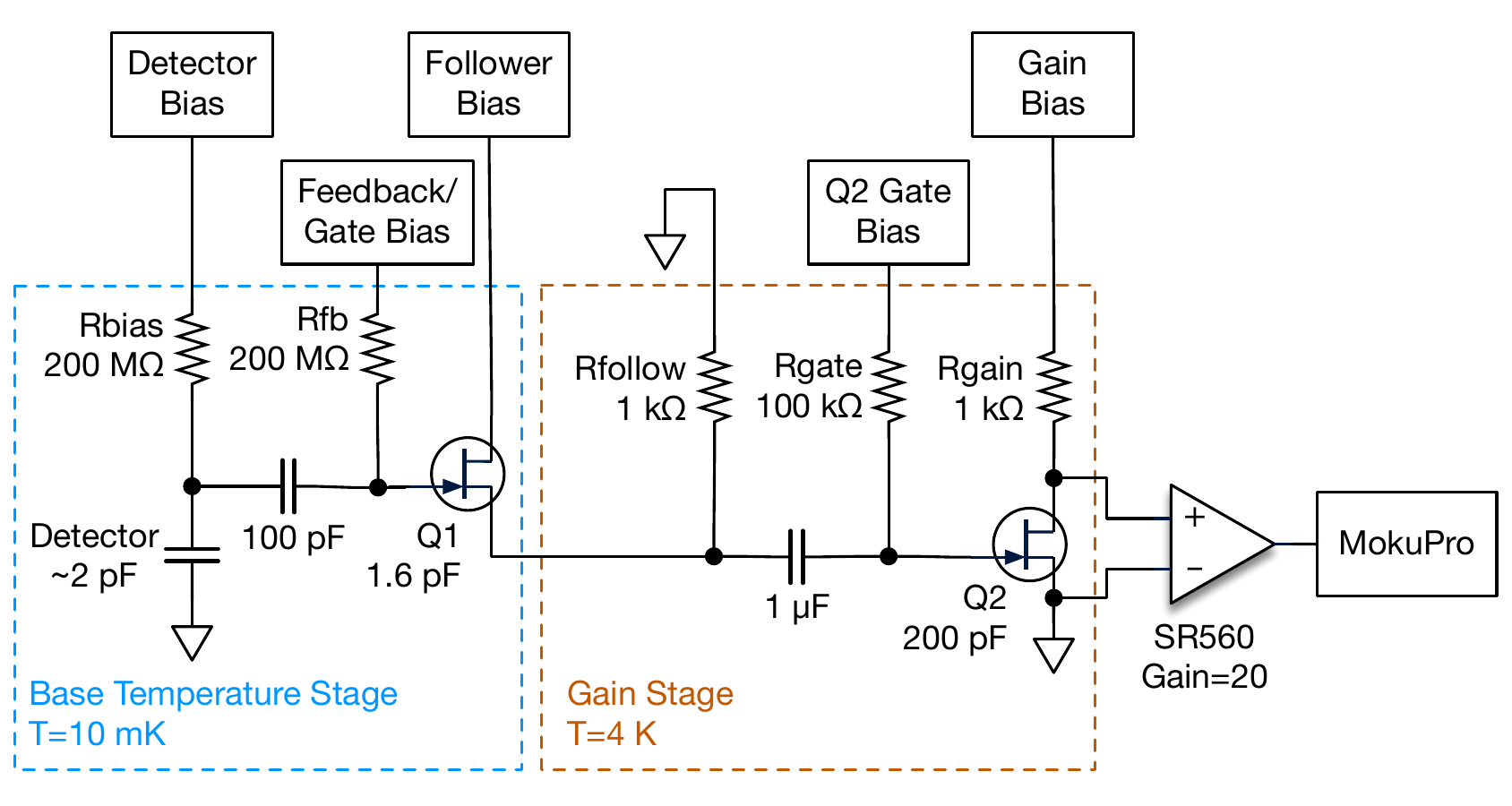}
    \label{fig:schematic}
    \caption{The simplified amplifier topology. Charge produced within the detector is integrated onto Q1, which acts as a voltage buffer. The signal is amplified at the 4K gain stage, and undergoes a further stage of room temperature amplification before digitization. The accrued charge on the Q1 gate is passively reset via the feedback resistor $R_\text{fb}$.}
\end{figure}
\section{Two-Stage HEMT Amplifier}
\begin{figure}
    \centering
    \includegraphics[width=\linewidth]{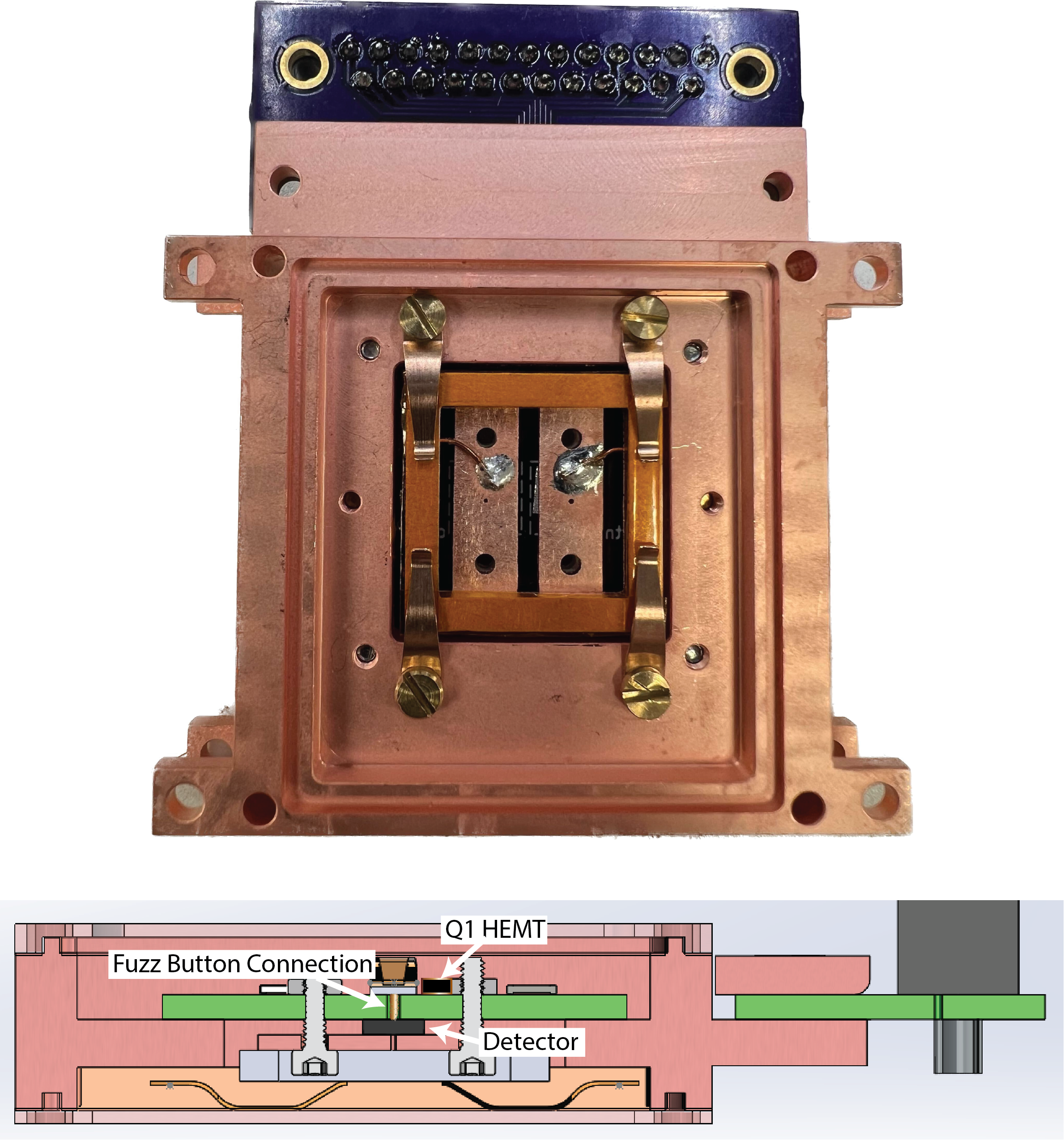}
    \label{fig:housing}
    \caption{\textbf{Top:} Picture of the backside of the prototype detector housing with silicon test devices installed. \textbf{Bottom:} Cutaway view of the side of the CAD model of the detector housing. Shown is the detector assembly highlighting the low capacitance connection between the detector and the front end HEMT of the two stage amplifier.}
    
\end{figure}

\emph{Design--}The approximate resolution scaling for a charge-integrating amplifier depends on the noise spectrum of the front end amplifier ($N_V$), the total capacitance of the system, the charge collection efficiency $\varepsilon_\text{CCE}$, and the inverse bandwidth $\tau$ as~\cite{Juillard_2019, PHIPPS2019181}
\begin{equation}
    \sigma_q \propto \frac{N_V}{\varepsilon_\text{CCE}\tau} \left(C_\text{in} + C_\text{det} + C_\text{par}\right),
    \label{eq:res}
\end{equation}
where the total capacitance has been written as a sum of the input capacitance of the amplifier $(C_\text{in}$), the capacitance of the detector ($C_\text{det}$), and the parasitic capacitance present at the amplifier input ($C_\text{par}$). To optimize the charge resolution, we use a two-stage amplifier topology (Fig. ~\ref{fig:schematic}) in which the front-end stage is integrated into the detector housing held at $10\,\mathrm{mK}$ and is configured as a ``common-drain'' voltage follower with approximately unity gain. This front-end HEMT has been specifically designed to reduce low frequency noise over the signal bandwidth ($\sim$10\,Hz -- 1\,MHz), minimizing $N_V$, and has a gate-to-source capacitance of $C_\text{in}=1.6\,\mathrm{pF}$. This two stage design further allows us to reduce the parasitic capacitance in two ways. First, the voltage-follower stage buffers the signal from upstream parasitic capacitance (e.g. the NbTi wiring to the higher temperature stages) and second, the distance between the the gate of the front-end HEMT and the detector contact is drastically reduced to $\mathcal{O}(\mathrm{mm})$ as shown in Fig.~\ref{fig:housing}. From initial calculations, we estimate this total parasitic capacitance to be at the level of $C_\text{par}\approx 1\,\mathrm{pF}$. 

The remaining term in our capacitive budget is the detector capacitance itself. Using a point contact to shape the E-field, detector capacitances of $\mathcal{O}(1\,\mathrm{pF})$ have commonly been achieved in large HPGe detectors~\cite{Cattadori_2011}. Further, we have measured the capacitance of the gram-scale devices synthesized by the SPLENDOR experiment to be on the order of a few pF. For purposes of bench-marking the charge performance of our charge amplifier, in this work we adopt the conservative estimate that the total capacitance seen at the input of the amplifier is $C_\text{total} = 5\,\mathrm{pF}$.

\emph{Readout--}The 10\,mK buffer stage capacitively couples to the detector and acts as a transimpedance amplifier. The high-value feedback and bias resistors integrate the drift current produced within the detector into a voltage signal at the gate of Q1. A NbTi twisted-pair cable connects the buffer stage to the 4K gain stage. The gain stage hosts the common-drain source resistor and couples the buffered voltage signal to the gate of Q2. The higher capacitance Q2 is configured as a common-source voltage amplifier, providing cryogenic gain. Via a low resistance copper cable, the 4K gain stage connects to room temperature electronics, which include a Stanford Research Systems SR 560 differential amplifier, low noise DC power supplies, and a Moku:Pro for data acquisition~\cite{watkins2023splendaq}. 

Both the base stage and gain stage utilize commercial low frequency, low noise cryoHEMTs~\cite{7021379}. The gain stage uses a 200 pF HEMT with an estimated 30-50 gain, while the buffer stage utilizes a 1.6 pF HEMT with a gain of approximately unity. Within the amplifier bandwidth, the total open loop gain is given by
\begin{equation}
    G_{OL} = \left(\frac{g_{m1}R_\text{follow}}{1+g_{m1}R_\text{follow}}\right)\left(-g_{m2}R_\text{gain}\right)
\end{equation}
where $g_{m1}$ is the transconductance of Q1 and $g_{m2}$ is the transconductance of Q2.

The base stage input HEMT (Q1) is coupled to the detector through a 100 pF (or larger) coupling capacitor, which is connected directly to the detector via a fuzz button as shown in Fig.~\ref{fig:housing}. We estimate that this compact geometry between the detector and the gate of Q1 should make $C_\text{par}\sim$1 pF.

\section{Voltage Noise Performance and Estimated Charge Resolution}
For the preliminary testing discussed in this work, a 100 k$\Omega$ feedback resistor was installed on the base temperature board and the amplifier was run without a detector. In this configuration, the resistor Johnson noise at $T=10\,\mathrm{mK}$ is negligible compared to the expected amplifier noise. While this relatively low resistor value prevents the input network from acting as an effective charge integrator, the standalone amplifier voltage noise performance can be measured. The amplifier was installed in an Oxford Proteox dilution refrigerator with a base temperature of 9.5 mK, including the amplifier power dissipation.

The voltage gain of the amplifier was found to be 36 at 10\,Hz (after removing the SR560 contribution) by injecting a known amplitude sine wave into the Q1 gate via the feedback resistor. The gain at higher frequencies was not measured due to the room temperature power supply filtering required for low noise performance. In separate tests without power supply filtering, the gain was found to be stable up to several hundred kHz with the bandwidth being determined by the time constant formed by the 4K gain resistor and output cable capacitance.  

\begin{figure}
    \centering
    \includegraphics[width=\linewidth]{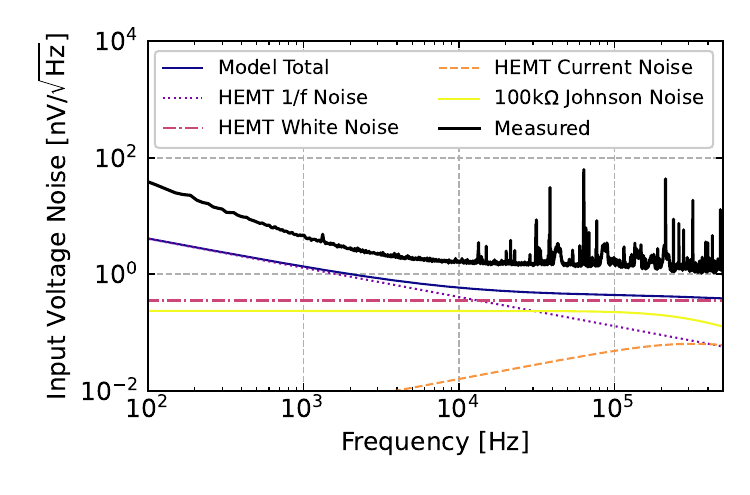}
     \label{fig:noise}
    \caption{The measured amplifier voltage noise spectrum, with the base temperature board at $T=10\,\mathrm{mK}$. The theoretical noise spectrum of the amplifier combines the HEMT white noise, current noise, and $1/f$ noise with 100k$\Omega$ feedback resistor. The high noise floor is dominated by room temperature electronics coupling down to the input of the amplifier.}
\end{figure}

The averaged input-referred noise voltage performance of the amplifier is shown in Fig.~\ref{fig:noise}, constructed from 300 individual 100\,ms traces sampled at 1\,MHz, divided by the measured gain of the cryogenic and room temperature amplifiers. Operating parameters of Q1 and Q2 are summarized in Table \ref{tab:q1q2}. Also shown are the expected noise contributions using a model based on Ref.~\cite{Juillard_2019}. The higher observed noise may be due to the lower operating power of the Q1 HEMT (1.3 $\mu$W), which are typically biased at 100 $\mu$W.

\begin{table*}
\caption{The Q1 and Q2 gate area, gate-source capacitance, gate-drain capacitance, operating temperature, drain-source voltage, drain-source current, and power dissipation.}
\begin{tabular}{cccccccc}
\toprule
HEMT & Area [$\mu$m$^2$] & C$_\text{gs}$ [pF] & C$_\text{gd}$ [pF] & T [K] & V$_\text{ds}$ [mV] & I$_\text{ds}$ [$\mu$A] & Power [{$\mu$}W] \\
\midrule
Q1 & $5.0\times$10$^2$ & 1.6 & 0.8 & 0.01 & 25 & 50 & 1.3\\
Q2 & 1.5$\times$10$^5$ & 236 & 8.9 & 4 & 89 & 940 & 83.7\\
\botrule
\end{tabular}
\label{tab:q1q2}
\end{table*}


To estimate the expected charge resolution, the measured amplifier noise spectrum is added in quadrature to the theoretical $T=10\,\mathrm{mK}$ Johnson noise of a 100 M$\Omega$ resistor (the parallel combination of the bias and feedback resistors). The total input capacitance ($C_\text{in}+C_\text{det}+C_\text{par}$) is assumed to be 5 pF. An estimated charge resolution of 7.2 electrons is calculated from using the combined noise spectrum and pulse template with the optimal filter formalism~\cite{PHIPPS2019181}.

\section{Conclusion}
We have developed a cryogenic split-stage amplifier for use with novel narrow bandgap semiconductor ionization detectors developed by the SPLENDOR experiment. We have shown that our low-capacitance HEMTs perform without issue at $10\,\mathrm{mK}$ and with low power dissipation. Combining our preliminary noise spectrum with an optimal filter calculation achieves a one-sigma estimated charge resolution of 7.2 electrons. The unique geometry of this low-noise charge amplifier allows it to be easily coupled to any insulating detector material---making it possible to generically study novel material substrates as detector prototypes with never-before-achieved charge resolution.

\section{Acknowledgments}

This work was supported by the U.S. Department of Energy through the Los Alamos National Laboratory. Los Alamos National Laboratory is operated by Triad National Security, LLC, for the National Nuclear Security Administration of U.S. Department of Energy (Contract No. 89233218CNA000001). Research presented in this article is supported by the Laboratory Directed Research and Development program of Los Alamos National Laboratory under project numbers 20220135DR, 20220252ER, 20230777PRD1, 20230782PRD1, and is receiving support from US DOE HEP Detector R\&D KA-25 program. This work was also supported in part by the US Department of Energy Early Career Research Program (ECRP) under FWP 100872. J. Anczarski is supported by the Kavli Institute for Particle Astrophysics and Cosmology Chabollah Fellowship. Support for A. T. J. Phipps was provided by a faculty support grant from the Cal State East Bay Division of Academic Affairs.


%

\end{document}

%% file: authors.tex
\author{J.~Anczarski}\email{anczarski@stanford.edu}\affiliation{Stanford University, Stanford, CA 94305, USA}\affiliation{SLAC National Accelerator Laboratory, Menlo Park, CA, 94025, USA}\affiliation{Kavli Institute for Particle Astrophysics and Cosmology, Stanford University, Stanford, CA, 94035, USA}
\author{M.~Dubovskov} \affiliation{Santa Clara University, Santa Clara, CA 95053, USA}
\author{C. W.~Fink}\email{cwfink@lanl.gov}\affiliation{Los Alamos National Laboratory, Los Alamos, NM 87545, USA}
\author{S.~Kevane}\affiliation{Stanford University, Stanford, CA 94305, USA}
\affiliation{SLAC National Accelerator Laboratory, Menlo Park, CA, 94025, USA}
\affiliation{Kavli Institute for Particle Astrophysics and Cosmology, Stanford University, Stanford, CA, 94035, USA}
\author{N. A.~Kurinsky}\affiliation{SLAC National Accelerator Laboratory, Menlo Park, CA, 94025, USA}
\affiliation{Kavli Institute for Particle Astrophysics and Cosmology, Stanford University, Stanford, CA, 94035, USA}
\author{A.~Mazumdar}\affiliation{Los Alamos National Laboratory, Los Alamos, NM 87545, USA}
\author{S. J.~Meijer}\affiliation{Los Alamos National Laboratory, Los Alamos, NM 87545, USA}
\author{A.~Phipps}\email{arran.phipps@csueastbay.edu}\affiliation{California State University, East Bay, Hayward CA 94542, USA}
\author{F.~Ronning}\affiliation{Los Alamos National Laboratory, Los Alamos, NM 87545, USA}
\author{I.~Rydstrom}\affiliation{Santa Clara University, Santa Clara, CA 95053, USA}
\author{A.~Simchony}\affiliation{Stanford University, Stanford, CA 94305, USA}
\affiliation{SLAC National Accelerator Laboratory, Menlo Park, CA, 94025, USA}
\affiliation{Kavli Institute for Particle Astrophysics and Cosmology, Stanford University, Stanford, CA, 94035, USA}
\author{Z.~Smith}\affiliation{Stanford University, Stanford, CA 94305, USA}
\affiliation{SLAC National Accelerator Laboratory, Menlo Park, CA, 94025, USA}
\affiliation{Kavli Institute for Particle Astrophysics and Cosmology, Stanford University, Stanford, CA, 94035, USA}
\author{S. M.~Thomas}\affiliation{Los Alamos National Laboratory, Los Alamos, NM 87545, USA}
\author{S. L.~Watkins}\affiliation{Los Alamos National Laboratory, Los Alamos, NM 87545, USA}
\author{B. A. Young}\affiliation{Santa Clara University, Santa Clara, CA 95053, USA}